\documentclass{pasj00}
\begin{document}
\SetRunningHead{S. Karino}{Light Curve of X-Ray Binary Pulsars}
\Received{2006/09/yy}
\Accepted{2007/xx/yy}

\title{Radiative Column and Light Curve of X-Ray Binary Pulsars}

\author{%
  Shigeyuki \textsc{Karino}\altaffilmark{1,2,3} }
 \altaffiltext{1}{JAD Program, Universiti Industri Selangor, Block 4, 
Jln Zirkon A7/A, \\
Seksyen 7,Shah Alam 40000, Selangor, Malaysia}
 \altaffiltext{2}{SISSA, ISAS, via Beirut 2/4, 34014 Trieste, Italy}
 \altaffiltext{3}{Shibaura Institute of Technology, Toyosu 3-7-5, Koto, Tokyo}
 \email{karino@sic.shibaura-it.ac.jp}


\maketitle

\begin{abstract}

We examine the published light curves (LCs) of 117 X-ray binary pulsars, 
focusing on the dependence of their light curves on the observed energy bands.
It is found that the energy dependence of the LCs appears only 
when the X-ray luminosity is larger than $\sim 5 \times 10^{36} \rm{erg~s}^{-1}$.
Assuming that the behavior of light curve is related to the radiative accretion 
column on the neutron star surface, this energy threshold can be considered 
as the observational proof of the accretion column formation proposed by 
Basko and Sunyaev.
Once we can grasp the existence of radiative column, we can also obtain 
several useful informations on the neutron star properties.
As an instance, we perform the statistical analysis of the orientation angle of 
the magnetic axis, and we find that the inclination angle of magnetic axis should 
be small in order to explain the observed statistics.

\end{abstract}

%------------------------------------------------------------
\section{\label{sec:1}Introduction}
%------------------------------------------------------------

After the discovery of the pulsating X-ray sources (Giacconi
et al. 1971; Tananbaum et al. 1972), a huge number of observational 
and theoretical studies of X-ray binary pulsars (XBPs) have been published.
The basic concept of X-ray pulsars has been understood
in a quite early stage of these studies; that is, 
the periodical X-ray pulsation is considered to be originated from 
accreting magnetized neutron stars
 (theoretically, Pringle \& Rees 1972; Lamb et al. 1973; 
observationally, 
White et al. 1983; Nagase 1989; Bildsten et al. 1997).

In the standard model of X-ray pulsars, the X-ray source is a rotating 
neutron star (NS) with a binary companion which feeds accretion matter 
via Roche lobe overflow or stellar wind.
If the NS has a strong magnetic field ($\gtrsim 10^{12} \rm{G}$), the
accreting matter eventually cannot accrete onto NS surface across 
the magnetic field lines.
Rather accreting matter is trapped to the magnetic field and constrained 
to follow the field lines, and near the stellar surface, it forms a funnel-like 
flow onto the magnetic poles 
(e.g. Pringle \& Rees 1972, Basko \& Sunyaev 1976). 
Consequently, the accreting matter is concentrated near the polar caps 
and emits high-energy radiation as the X-ray photons.  
Since generally the magnetic axis and the rotational axis are misaligned, 
these X-ray photons from the poles are observed as periodic pulsations 
(Shapiro \& Teukolsky 1983; White et al. 1983; Nagase 1989;  
Bildsten et al. 1997).

The basic model of X-ray pulsars has already been obtained.
Although the detailed investigations of X-ray sources in these 
systems are still difficult since the observational method of XBPs 
is limited in space experiments, the sample number of XBPs 
increase significantly in these days, thanks to recent development 
of space experiments.
Hence, even though what we can know from a single observation is less, 
we can distill some hints of X-ray source properties by statistical analysis 
of observed X-ray pulsars (Bulik et al. 2003; Raguzova \& Popov 2005).
For instance, Bulik et al. (2003) use the catalogue of XBP and propose
the new open question about the distribution of peaks of X-ray LCs 
concerning with the formation scenario of the NS binary systems.
They have counted the number of single and double peak X-ray LCs 
in published catalogues, and found that the observed number of single peak 
pulsars is relatively high.
If the inclination of magnetic axis from the rotational axis is random, 
the majority of observed LC should have double peaks.
Then, they argued that, in order to explain the large number of single 
peak LCs, the inclination angle of the magnetic axis with respect to 
the rotational axis must be small.

In the investigation of the NS physics, LCs of XBPs give 
us useful informations of the systems.
The unique strong point of LC observation is that it does not depend on 
the distance to the source and carries us original information of central NS.
Generally, the light curves of those XBPs are various.
Some of them show simple single/double peak pulses persistently, while
a lot of systems shows complicated light curves.
These complexity is caused by absorption by circumstellar matter, 
misalignment of the magnetic axis and rotation axis, Doppler effect, and so on.
Also the shape of the LCs often depend on the observed 
energy band (White et al. 1983; Nagase 1989).
Several mechanisms have been 
proposed in order to explain the energy dependency of LCs.
For example, if the matter surrounding the NS has absorption rate 
which depends of the energy band, the observed  X-ray LC will be 
different in each energy band (White et al. 1983; Marsden et al. 1998).
However, the real situation is still under discussion.

In this paper, we examine whether we can say something new about NS 
physics from recently provided XBP catalogues.
We mainly focus on the LC of XBP, which brings us 
the distance-independent information about the magnetized NS.
Firstly, we check the relationship of LC shapes with other observational
properties of XBP.
We find there is clear correlation between the LC complexity and 
the X-ray luminosity of XBP.
This correlation can be explained by classical radiation column model 
given by Basko, Sunyaev (1976). 
Then we make statistical analysis of the LC of XBP.
Statistics of the LC distribution can be considered to reflect the axis 
inclination conditions of central NS.
Our analysis suggests that the inclination angle between the magnetic axis 
and the rotational axis of the central magnetized NS should be small.
Also it indicates that the opening angles of the X-ray beams from 
the radiative column should be narrow.
Since these NS properties are difficult to obtain from direct 
observations, such a statistical investigation becomes 
important.

In the next section, we introduce the X-ray binary pulsar samples briefly. 
In Sec.~\ref{sec:3}, we remind the basic model of radiative zone of NSs.
In Sec.~\ref{sec:4}, the data and analysis of X-ray pulsar samples is 
shown.
In Sec.~\ref{sec:5}, we analyze the XBP sample statistically and discuss 
the obtained nature of central magnetized NS. 
The last section is devoted to the summary.

%------------------------------------------------------------
\section{\label{sec:2}Analysed systems of XBP}
%------------------------------------------------------------

Thanks to several space observations by ASCA, Beppo SAX, CGRO, 
Chandra, Ginga, INTEGRAL, ROSAT, XMM-Newton and so on, now we 
have a large number of XBP samples and candidates.
Firstly, we should mention that we are interested in the X-ray pulsars 
whose LCs are published. 
These XBPs are listed in several catalogues; here we refer to the 
catalogue of Be/X binaries and candidates given by 
Raguzova, Popov (2005), 
the catalogue of high-mass X-ray binaries given by Liu, van Paradijs, 
van den Heuvel (2000), the review and tables of discrete X-ray sources 
in Small Magellanic Cloud (SMC) given by Yokogawa et al. (2003), 
the table of Be/X-ray binaries in SMC given by Haberl, 
Pietsch (2004), the 2nd IBIS/ISGRI soft gamma-ray survey catalogue 
given by Bird et al. (2005), and the list of observed light curves
of X-ray pulsars given by Bulik et al. (2003).
From these previous works, we chose 117 X-ray pulsars whose spin 
periods are known.
70 of them are Galactic sources, 41 are in SMC and 6 are LMC sources.
These chosen systems are listed in Tables.~\ref{tab:listG} -- \ref{tab:listL}.
Even comparing with the generic and comprehensive previous analysis  
of X-ray pulsars given by White et al. (1983), Nagase (1989), Bildsten et al. 
(1997), Bulik et al. (2003) and so on, our sample number is the largest.

We prepare the following three tables; systems in Galaxy, 
Small Magellanic Cloud (SMC), and Large Magellanic Cloud (LMC).
In these tables, the X-ray binary systems are sorted by the 
spin period. 
In the third column of the tables, we show the type of light curves, 
which is categorized as following.
Among all the sample XBP, for 90 systems, the LCs have been taken 
and published in at least 2 energy bands.
Then we categorize those pulsars into 2 groups: Regular LC group and 
Irregular LC group.
The diagnosis is the following.
When we compare two light curves obtained in different energy bands,
if (i) the number of peaks are different, and/or (ii) the phase of the peak
is shifted (roughly more than 20 \%), then we categorize this pulsar as 
Irregular group~\footnote{For dim sources and sources without enough 
statistics, small peaks of LCs could be missed. Hence there may be a 
tendency that dim XBPs are easy to be counted as Regular group.
On the other hand, the diagnosis of the peak shift has no bias with 
regard to the luminosity.}.
Otherwise, it is put into Regular group.
Some pulsars show marginal behaviors of LC, hence firstly we 
categorize them into Regular group but indicate them as "Rm" 
in the Tables. 
For the systems whose light curve is published only for one energy band, 
we can not make diagnosis and indicate ``---'' in this column.

The fourth and fifth columns are devoted to the spin period of the NS 
and the orbital period of the system, respectively.
The spin period $P_{\rm s}$ is shown in unit of second, while the orbital 
period $P_{\rm o}$ is in unit of day.
In the sixth column, the X-ray luminosity $L_{\rm X}$ ($\rm{erg~s}^{-1}$) 
is shown.
Note that many systems of XBP are observed as transient X-ray sources.
For almost all the systems, $L_{\rm X}$ coincides to the maximum 
luminosity of the system, and their LCs are taken in this bright phase.
The surface magnetic field strength of the NS $B_{\rm p}$, 
which can be estimated from cyclotron line, is shown in seventh column.
In the final column, we show the additional information.
Also the references of light curves are indicated here.
We have to keep in mind that those data has been taken by different 
experiments, and hence the total luminosity of these XBPs are estimated 
by different energy bands. 
This difference may make uncertainty of the total luminosity shown in 
the tables.

%------------------------------------------------------------
\section{\label{sec:3}Radiative column}
%------------------------------------------------------------

The luminosity and mass accretion rate is strongly connected in NS
accreting systems.
In Basko, Sunyaev (1976), they derived the "accretion rate" as
\begin{equation}
L_{\rm{t}} \equiv \dot{M} \frac{GM}{R} ,
\end{equation}
where $\dot{M}$ is mass-accretion rate.
And they proposed a critical accretion rate which is described as
\begin{eqnarray}
L^{\ast} &=& \frac{2 c l_0}{\kappa} \frac{GM}{R} \nonumber \\
&=& 4 \times 10^{36} \left( \frac{\sigma_{\rm{T}}}{\sigma_{\rm{s}}} \right)
\left( \frac{l_0}{2 \times 10^5 \rm{cm}} \right) 
\left( \frac{10^6 \rm{cm}}{R} \right)  
\left( \frac{M}{M_{\odot}} \right)
 \rm{erg~s}^{-1} \nonumber \\
&=& 6 \times 10^{36} \left( \frac{\sigma_{\rm{T}}}{\sigma_{\rm{s}}} \right)
\left( \frac{l_0}{2 \times 10^5 \rm{cm}} \right)  \rm{erg~s}^{-1} ,
\end{eqnarray}
for a canonical NS with $R = 10^6 [\rm{cm}]$ and $M = 1.5 M_{\odot}$.
In the first equation, $\kappa$ denotes the opacity.
Here, $\sigma_{\rm{s}}$ denotes the isotropic 
scattering cross-section, and $\sigma_{\rm{T}}$ is the Thomson 
cross-section, respectively.
$l_0$ indicates the circumference of the accretion cylinder (see, Fig.~1 in
Basko, Sunyaev 1976).
They predict that if $L_{\rm{t}}$ is smaller than $L^{\ast}$, then 
the accretion channel is wholly optically thin.  
Hence, the X-ray radiation is mainly emitted from the small cross section 
of the accretion column on the surface and the main beam is emitted into 
the direction along to magnetic axis.
On the other hand, if $L_{\rm{t}}$ is larger than $L^{\ast}$, the optical 
depth of the accretion column above the NS surface will be large.
This optically thick region is called radiative column.
In this radiative column, the radiative pressure is no longer negligible and 
the accreting fluid decelerates significantly.
Passing this radiative column, the fluid converts its kinetic 
energy into radiative energy.
Since the emitted photons cannot go straight in this dense region, 
the photons are scattered and diffusively go out from side face of the column.
Here, note that, since the geometry of the accretion flow is not spherical,
the radiation pressure becomes important even when the luminosity is less 
than the Eddington limit for spherical accretions.

The emission manner depends on the existence of accretion 
column.
If the opaque radiative column appears, the X-ray radiation will be emitted 
from the side-face of the column, while the radiation is emitted in the polar 
direction from the cross-section of the column when such a column does 
not exist.
X-ray emission in the direction of magnetic pole is called the 
{\it pencil-beam}, while side-face radiation is called the {\it fan-beam}. 
Considering these two manners of X-ray radiation,  we can expect 
that the critical accretion rate $L^{\ast}$ can be understood also as 
the threshold of exchange between these two types of radiation.
Actually, there are examples which shows the dependence of the LC 
on the X-ray luminosity.
For instance, the time dependent LC and $L_{\rm X}$ of EXO2030+375 
(G28) has been taken and the results show that the phase of the primary 
peak is completely changed with the threshold of 
$L_{\rm X} \approx 10^{37} [{\rm erg~s}^{-1}]$ (Parmar et al. 1989a,b) 
This shows the configuration of the accreting column changes at this 
threshold luminosity and subsequently the beam type is changed 
(Nagase 1989).

%------------------------------------------------------------
\section{\label{sec:4}Data analysis of XBPs}
%------------------------------------------------------------

\begin{figure}
  \centering
	\FigureFile(80mm,60mm){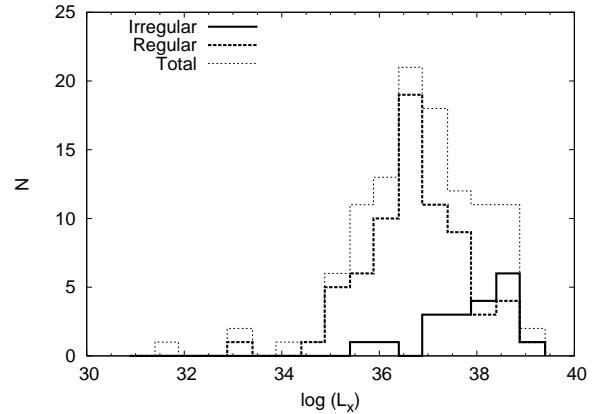}
  \caption{
    Number counts of Regular and Irregular pulsar luminosities.
    The luminosity distributions of Irregular group, Regular group, 
    and total sample including pulsars whose LC is unpublished, are shown
    by thick solid, thick dashed, and thin dotted lines, respectively.
    From this figure, we can see that the distribution of Regular and 
    Irregular groups are different.
    Two exceptionally dim Irregular pulsars are G30 (XTEJ1906+09) 
    and S28 (RXJ0059.3-7223), which are considered to be wind-fed type.
	\label{fig:n-lx}
  }	
\end{figure}

First of all, we check whether the observational quantities of XBP show 
any difference between Irregular and Regular groups.
In Fig.~\ref{fig:n-lx}, the luminosity distributions of Irregular group, 
Regular group, and the whole sample of  XBP are shown by thick 
solid, thick dashed, and thin dotted lines, respectively.
From this figure, we can see that Regular and Irregular groups have 
different distributions (with more than 99\% probability according 
to the Kolmogorov-Smirnov test).
The Regular group shows almost symmetric distribution with an axis 
of symmetry around 
$L_{\rm{X}} \approx 2 \times 10^{36} \rm{[erg~s^{-1}}]$, 
and the upper and lower tails spread broadly.
On the other hand, Irregular group distributes in a narrow region
above $L_{\rm{X}} > 5 \times 10^{36} \rm{[erg~s^{-1}]}$.
Note that there are two exceptionally dim Irregular LC pulsars, 
and they are G30 (XTEJ1906+09) and S28 (RXJ0059.3-7223).
Both of them have giant companions and are fed by wind accretion.

Here we note that, especially for Galactic sources, it is difficult to 
estimate the distance of the source precisely.
Because of this uncertainty, the luminosities of Galactic XBPs may 
have errors as large as factor of 10 (Haberl \& Sasaki 2000). 
On the other hand, the distances (and, hence, luminosities) of 
SMC/LMC XBPs can be estimate much precisely.
In order to avoid the uncertainties of the luminosities of Galactic XBPs, 
we can check the number counts of Regular and Irregular XBP 
luminosities only for SMC/LMC sources, and confirm that the resultant 
distribution is almost the same with Fig.~\ref{fig:n-lx}.
Also we would like to comment that the uncertainties of the luminosity 
could broaden the distributions not only of Regular group but also of 
Irregular group in Fig.~\ref{fig:n-lx}. 
Consequently the inconsistency of the distributions will be still essential,
even with the uncertainty.

%-------------------------------------------------------------
%------------------------------------------------------------

\begin{figure}
  \centering
	\FigureFile(80mm,60mm){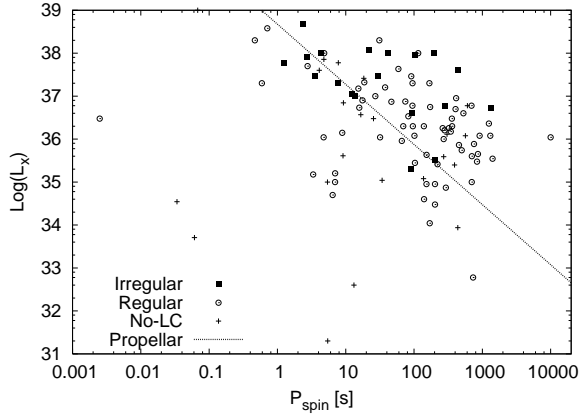}
  \caption{
    Spin period distribution regarding to luminosity is shown
    for Irregular (star) and Regular (cross) LC pulsars.
    For comparison, No-LC group pulsars are also 
    shown (plus).
    Almost every pulsar distributes on upper-right region, while a few 
    exists in lower-left region (they are in the propeller regime, 
    or in the quiescent phase).
    The critical line of the propeller transition, for $\mu = 10^{30}$
    is also shown.
    It indicates that in the propeller regime, the LC must be regular.
  }
  \label{fig:ps-lx}
\end{figure}

In Fig.~\ref{fig:ps-lx}, the relationship between the spin period 
$P_{\rm{spin}}$ and the luminosity is shown for Regular (open-circle) 
and Irregular (filled-box) LC pulsars. 
For the purpose of the comparison, also No-LC group pulsars are shown 
by small  cross in the same figure.
Almost every pulsar distributes on upper-right region, while a few 
exists in lower-left region.
These exceptional points are corresponding to the pulsars which are 
in the quiescent phase or in the propeller regime (Bildsten et al. 1998).
According to the analysis given by  Ikhsanov (2003) and 
Raguzova, Popov (2005), the accretion mode exchanges from 
the accretion to the propeller regime on the basis of a certain criterion.
That is, if the condition
\begin{equation}
P_{\rm{spin}} < P_{\rm{prop}} = 81.5 \mu_{30}^{16/21} L_{36}^{-5/7}
\label{eq:prop}
\end{equation}
is satisfied, the propeller effect does work, and accretion will be 
prevented.
Here, $\mu_{30}$ and $L_{36}$ are the normalized magnetic moment and 
luminosity, respectively. 
At a glance, the Irregular pulsars distribute almost in the same region as
Regular pulsars.
Notice that, however, entire Irregular group locates above, or near, 
the critical line of the propeller transition, shown by Eq.~(\ref{eq:prop}).
It indicates that in the propeller regime, the LC becomes regular, and 
it is consistent with the argument of Basko, Sunyaev (1976).
Namely, in propeller regime, the accretion rate must be much smaller, 
since the accreting matter will be blown away.
In this regime, and with consequent small accretion rate, it seems 
difficult to form the radiative dense column.
Hence, in this case, the X-ray radiation is mainly emitted as a 
pencil beam from tiny region on the stellar surface.
Then, in such regime, we can only observe simple pulse emitted in pencil
manner.

%-------------------------------------------------------------
%------------------------------------------------------------

\begin{figure}
  \centering
	\FigureFile(80mm,60mm){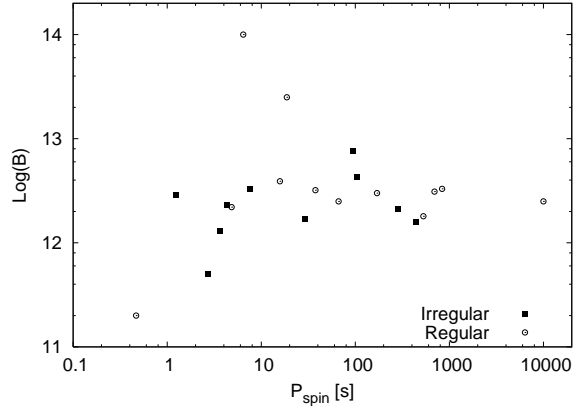}
  \caption{
    Spin period of Irregular and Regular LC pulsars are plotted 
    as the function of the magnetic field strength on the stellar surface.
    No significant difference can be seen between Irregular and Regular 
    groups.
    The exceptionally strong magnetic field pulsar with $\sim 10^{14}[{\rm G}]$ 
    is G14 (1E1048.1-5937) and it is perhaps an AXP.
    We may see weak dependence of spin period on the magnetic field.
  }
  \label{fig:ps-bp}
\end{figure}

Fig.~\ref{fig:ps-bp} shows the spin period of Regular and Irregular 
LC pulsars plotted regarding to the magnetic field strength 
on the stellar surface.
No significant difference can be seen between Regular and Irregular 
groups.
Since X-ray pulsar systems are not equilibrium systems, they are 
considered to be still in spin-evolutionary stage.
It means that the spin period needs not any relation with magnetic field.
There is one point which has exceptionally strong magnetic field 
($ \sim 10^{14} [{\rm G}]$), and this is G14 (1E1048.1-5937).
In our sample, G13, G14, G68 and S9 are the candidates of anomalous 
X-ray pulsars (AXPs).
If they are, the radiative mechanism could be different from XBPs, 
hence we should remove them from our sampling.
In the analysis, we do not have the necessary LCs for G13 and S9, 
and those pulsars are not included in the statistical results.
G68 is a reliable candidate of AXP, hence we exclude this from 
all analysis.
On the other hand, G14 is counted as a sample, hence this one 
point can be an error.
Historically, G14 has been considered as a candidate of XBP in the 
early stage (Seward et al. 1986), however, it is supposed as 
a candidate of an AXP in the last few years (Tiengo et al. 2002).

%-------------------------------------------------------------
%------------------------------------------------------------

\begin{figure}
  \centering
	\FigureFile(80mm,60mm){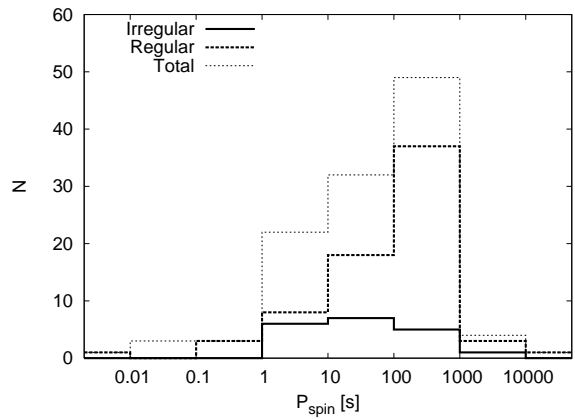}
  \caption{
    Number counts of spin periods of Irregular and Regular pulsars.
    The spin period distributions of Irregular group, Regular group, 
    and total sample including pulsars whose LC is unpublished, are shown
    by thick solid, thick dashed, and thin dotted lines, respectively.
  }
  \label{fig:n-ps}
\end{figure}

In Fig.~\ref{fig:n-ps}, the spin period distributions of Regular group, 
Irregular group, and total number including pulsars whose LC is 
unpublished, are shown by thick solid, dashed, and dotted lines, respectively.
The number of observed Regular pulsars increases with spin period and 
truncates at around 1,000 sec, while, the number of Irregular group shows 
rather flat distribution.
The spin of the magnetized NS accelerates via accretion process and 
decelerates in propeller regime.
The Irregular group pulsars, however, locate in the parameter 
region where propeller effect does not work (see, Fig.~\ref{fig:ps-lx}).
Hence, those Irregular pulsars do not suffer from the spin down mechanism.
Then, they may keep their spin rapid and show rather flat distribution 
as shown in Fig.~\ref{fig:n-ps}.

%-------------------------------------------------------------
%------------------------------------------------------------

\begin{figure}
  \centering
	\FigureFile(80mm,60mm){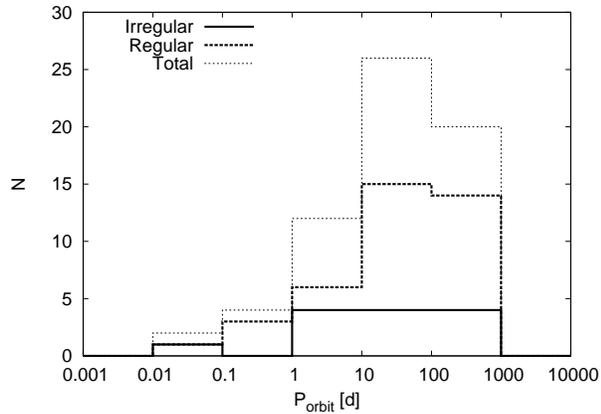}
  \caption{
    Number counts of orbital periods of Irregular and Regular pulsars.
    Lines are the same as in Fig.~\ref{fig:n-ps}.
    The orbital period number-counts of Irregular group shows quite flat 
    distribution, while Regular group has a peak around $\sim 100 [{\rm d}]$.
  }
  \label{fig:n-po}
\end{figure}

In Fig.~\ref{fig:n-po}, the orbital period distributions of Regular group, 
Irregular group, and the total number including pulsars which LC is 
unpublished, are shown by solid, dashed, and dotted lines, respectively.
From this figure, it is clear that Irregular group shows quite flat 
distribution of the orbital period, while Regular group has somehow 
a peak around 100 d.
Unfortunately, it is difficult to observe long orbital periods, say, 
longer than 1,000 d, since they need a long observational time 
(several years).
Hence, near the upper end, we may have observational bias.
Here, we may expect the following scenario: the longer orbital 
period means the larger orbital separation. 
The larger separation may cause smaller accretion rate, if the mass of 
the primaries are almost the same.
Hence, from the condition given by $L_{\rm{t}} < L^{\ast}$, the number of 
Irregular group decreases in large $P_{\rm{orbit}}$ region.
Although this expectation matches to the tendency of the figure, 
it is required better statistics to confirm it.

%-------------------------------------------------------------
%------------------------------------------------------------

%------------------------------------------------------------
\section{\label{sec:5}Discussions}
%------------------------------------------------------------

As shown in Fig.~\ref{fig:n-lx}, Irregular group has clear minimum X-ray 
luminosities around $L_{\rm{X}} \approx 5 \times 10^{36} \rm{erg~s}^{-1}$, 
and this minimum limit coincides with the criterion of radiative column 
given by Basko, Sunyaev (1976).
Namely, it is natural to consider that the {\it irregularity} of LC correlates 
with the existence of radiative column with finite height.

Although there are two exceptions in Fig.~\ref{fig:n-lx}, these two XBPs 
have different type of companions and different accretion manners: 
wind-fed accretions from large companions.
That is, the central NS of these systems are surrounded by huge amount 
of circumstellar matter.
Because of the large absorption by the surrounding matter, the small peak 
may be screened.
Also the clumpiness of the surrounding matter may cause 
the phase-dependent concealments of LCs (Nagase 1989).
These effects will be significant in lower energy bands, and they lead to  
the energy-dependent LC shapes. 
Therefore, it may be possible to conclude that the irregularity of these 
pulsars is caused not by the configuration of radiative column, 
but by absorptions due to surrounding matter.
If this is the case, all the other Irregular pulsars satisfy the 
Basko-Sunyaev criterion, and can be considered to have radiative 
column with finite height.

Examining strictly, we have to treat the X-ray luminosity carefully.
The X-ray luminosities taken by different experiments are estimated 
based on different energy bands.
For XBP, it is difficult to estimate the total luminosity 
accurately, only from fragments of the spectra.  
However, in our sample, almost Irregular XBPs are observed in 
relatively broad energy bands, from 0.1 keV to several tens keV.
It means that the X-ray luminosities of Irregular XBPs are estimated 
based on large part of the total spectra, with relatively high accuracy.
Hence, our conclusion that the Irregular group has large X-ray 
luminosity is unchanged in our sample, even under the consideration of 
the uncertainty of the luminosity due to the lack of the spectra.

Reminded that, if a NS does not have a radiative column, 
X-rays will be emitted only as pencil beams in our basic conjecture.
Otherwise, X-ray LC is affected from the contribution of fan-beams. 
Next we check that the radiative column really affects the shape of LC.
And then, we examine whether we can obtain further informations of 
magnetized NS from the statistical analysis of LC.

\subsection{Light curve models}

The X-rays from a NS are radiated in two different manners, that is, 
{\it pencil beam} and {\it fan beam}.
The intensities of these two beams have the angular dependence; 
the pencil beam gives the maximum intensity when the line of sight 
coincides with the magnetic axis, while the fan beam gives the maximum 
intensity when we observe the NS from the magnetic equator.
Hence, as the simplified models of beams, we assume that the intensity
(number of photons) can be described as
\begin{eqnarray}
F (\alpha, \psi) & \propto & 
\left\{
\begin{array}{cc}
\cos^{\mu} \alpha & {\rm{(pencil ~beam)}}  \\ 
\sin^{\nu} \alpha &  {\rm{(fan ~beam)}} ,
\end{array}\right.
\label{eq:beam}
\end{eqnarray}
where $\alpha$ denotes the angle between the line of sight and 
the magnetic axis.
$\mu$ and $\nu$ denote certain factors generally depending on the spin 
phase $\psi$.
This $cos$ -- $sin$ dependence comes from the radiating area seen 
by the observer (Nagel 1981).
We assume simple dipole magnetic poles, and do not include the effect of 
screening by the stellar surface, hence the both beams are observed at 
$- \pi / 2 \leq \alpha \leq \pi / 2$.

\begin{figure}
  \centering
	\FigureFile(40mm,30mm){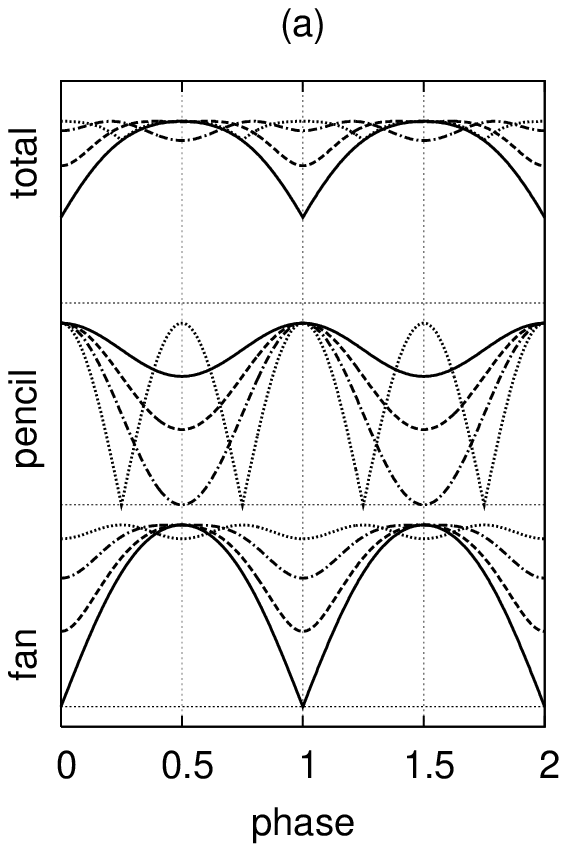}
	\FigureFile(40mm,30mm){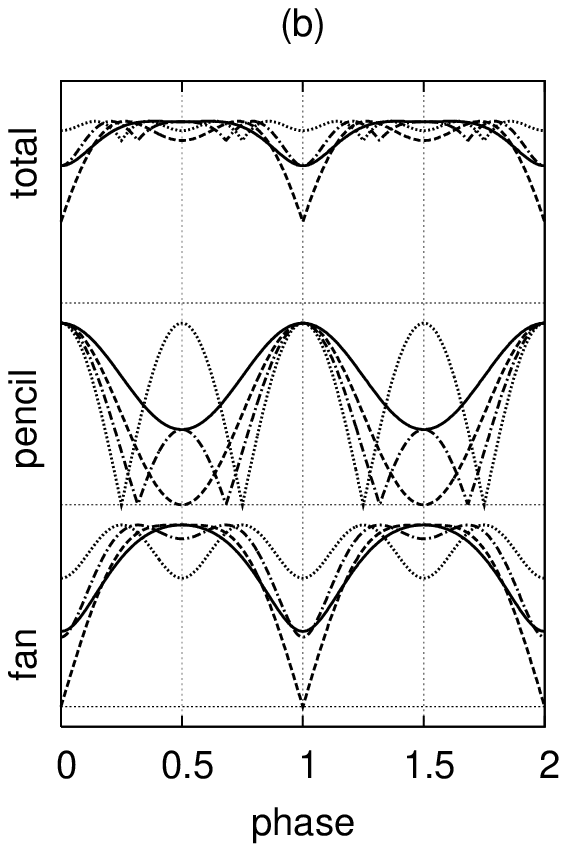}
	\FigureFile(40mm,30mm){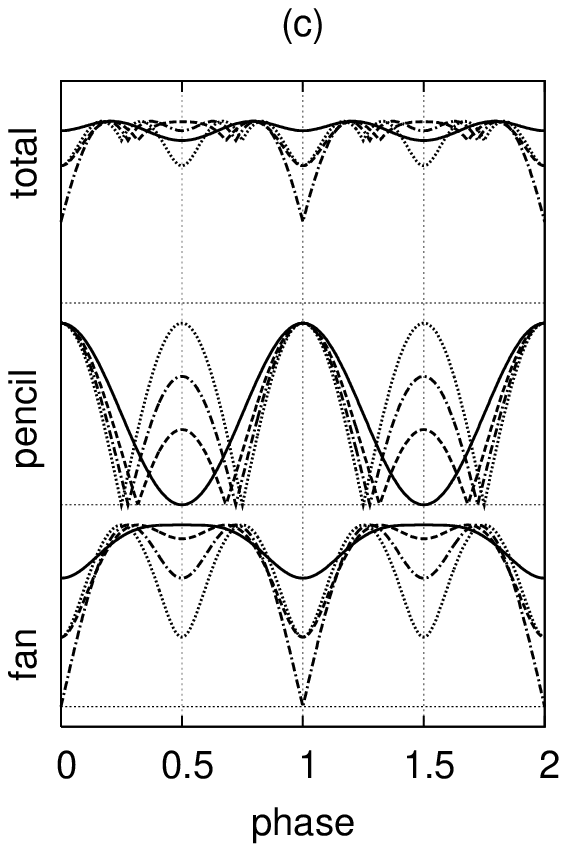}
	\FigureFile(40mm,30mm){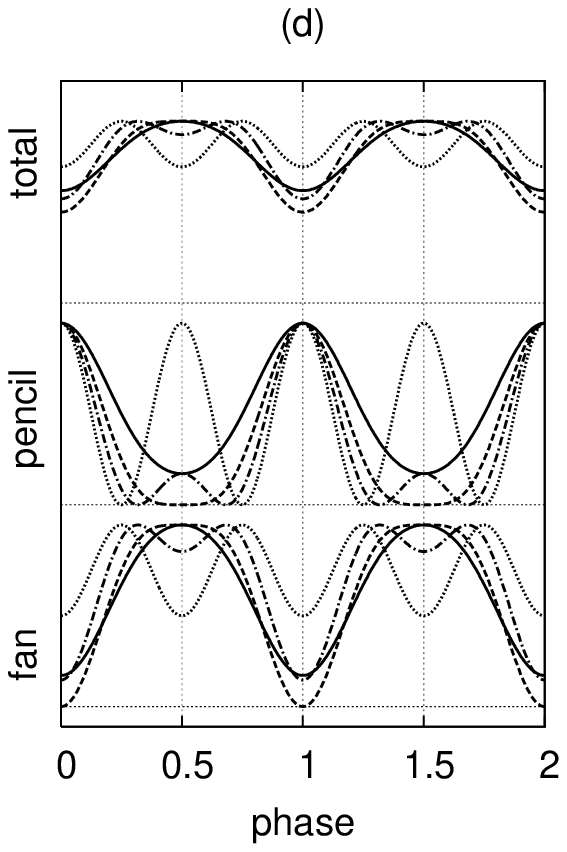}
  \caption{
    Model light curves. Pure-fan, pure-pencil, and combined beams 
    are shown in the branches entitled fan, pencil and total in the figure.
    The computed parameters are the following.
    Models (a), (b) and (c) are generated for 
    $\theta_{\rm{m}} = \pi / 8, \pi /4$ and $3 \pi / 8$, respectively.
    For these three cases, the mixing factor is fixed as $A = 0.5$ and 
    also we fix $\mu = \nu = 1$.
    Through all the figures, solid, dashed, dash-dotted, and dotted lines 
    correspond to $\theta_{\rm{o}} = \pi / 8, \pi /4, 3 \pi / 8$, and 
    $\pi / 2$, respectively.
    In model (d), we show the result for $\mu = \nu = 2$ but other values
    are the same as model (b).
     }
  \label{fig:lc1}
\end{figure}

In general, X-ray photons with only higher energies can be emitted 
as the pencil beam, since along the magnetic axis the mean free path 
becomes so short that weak radiation can not escape in this sense, 
when the accretion column exists.
On the other hand, from the side face of the accretion column, photons 
scattered and diffused many times are emitted as the fan beam.
These side-way photons can have lower energies in average.
When the height of the column is significant, the fan beam will be 
dominant, however, with a moderate height of column, pencil beam 
could not be neglected. 
Hence, the observed beams of X-ray from 
the accretion column will be the mixture of the fan beam and 
the pencil beam.
We model such a mixed beam as following;.
\begin{equation}
F (\alpha, \psi) = A \cos^{\mu} \alpha + B \sin^{\nu} \alpha  ,
\end{equation}
where $A$ and $B$ are mixing factors which are defined by $A + B = 1$ and 
$A \in [0,1]$.
Of course, $A=0$ corresponds to the fan-beam, while $B=0$ results in 
the pencil-beam.
In Fig.~\ref{fig:lc1}, we show some examples of the light curves generated in 
this manner, with different $\theta_{\rm{m}}$ and $\theta_{\rm{o}}$.
Here, $\theta_{\rm{m}}$ is the inclination angle between the magnetic axis 
and the rotational axis, while $\theta_{\rm{o}}$ is that between the rotational
axis and the line of sight.

If the radiative region has finite side-face, it is natural to consider that 
both beams are emitted in some amounts (mixed-beam).
When we observe the mixture of them, the mixed beam will show more 
complicated pulse shapes.
If we admit that the ratio $A/B$ depends on the energy of photons,
the shape of the LC will be changed for each energy band, and
the energy dependence of X-ray LCs can be explained naturally.
For instance, if we see Fig.~\ref{fig:lc1} , and compare pencil-, fan- and 
mixed-beams, it is easily seen that some of them show the phase inversion
and peak splitting.
Such a metamorphosis of the LC can be observed only when the radiative 
column exists. 
Hence, we may conclude that the existence of the radiative column can 
cause the irregularity of some XBPs.
Though it is difficult to explain all the irregularity of XBP LCs by only this 
mixed-beam hypothesis, some of the observed complicated LCs may be 
reconstructed by adopting this method.

%-------------------------------------------------------------
%------------------------------------------------------------

\subsection{Light curve statistics \label{ssec:lcs}}

Now, we have a large sample of pulsars listed in Table~1 -- 3.
According to Basko, Sunyaev (1976) and the previous discussion in  
this paper, when the X-ray luminosity of XBP is larger than 
$L^{\ast} \sim 5 \times 10^{36} [\rm{erg~s}^{-1}]$, 
the radiative column with finite height will appear on the NS surface.
If such a radiative column with finite height emits both the pencil beam and 
fan beam and we can observe both beams, the observed light curve may 
show an irregular behavior.
In our sample, however, not all the pulsars with $L_{\rm X} > L^{\ast}$ 
are categorized as Irregular group, and a large fraction of bright pulsars 
show Regular LC.
In fact we have 40 bright pulsars with 
$L_{\rm X} > 5 \times 10^{36} [\rm{erg~s}^{-1}]$ and 
24 of them are Regular members.
If all the bright pulsars should show the irregular behavior, 
this fraction is inconsistent.

Actually, even if the radiative column appears and both beams are emitted,
it does not mean every pulsar will be observed as irregular one.
For instance, when the magnetic axis and the line of sight almost 
coincide, the pulse component due to fan beam will be very weak. 
Hence, the observed LC will be dominated by pencil beam only, 
and cannot be irregular.
Also when the magnetic axis almost coincides with rotation axis, 
the irregularity will not appear since the LC becomes flat.
For another example, if the beam is collimated in small angle due to certain 
reasons (it corresponds to large $\mu$ and $\nu$ in Eq.~(\ref{eq:beam})), 
we may less likely observe both beams in different phase.
Also in this case, the observed LC will be dominated by only one of 
two beams and behave regularly.
Then, the possibility of observation of Irregular pulsars depends on the 
inclination angles and the opening angles of the beams.

Suppose that the opening angles of the pencil- and fan-beam are 
$\varphi_{\rm{p}}$ and $\varphi_{\rm{f}}$, respectively. 
When the inclination condition
\begin{equation}
2 \theta_{\rm{m}} + \varphi_{\rm{f}} + \varphi_{\rm{p}} > \frac{\pi}{2}
\label{eq:incl}
\end{equation}
is satisfied, there is a region where both the pencil beam and the fan beam 
can be observed from the same position.
The probability that the observer's location enters in this region can be 
described as
\begin{equation}
p_{\rm{ir}} = \frac{2 \gamma}{\pi} ,
\end{equation}
where $\gamma = 2 \theta_{\rm{m}} + \varphi_{\rm{f}} + \varphi_{\rm{p}} 
- \pi/2$.
$p_{\rm{ir}}$ denotes the probability that the observer can 
see both pencil- and fan-beam, and in this case the light curve will be 
categorized as Irregular group.
Of course, the probability that the light curve is observed as a Regular 
member is $p_{\rm{r}} = 1 - p_{\rm{ir}}$.
Note that though the beam opening angle depends on the strong magnetic 
field, the gravitational bending, beaming, etc., here $\varphi$ denotes the 
observable opening angle and contains all those effects.

Next, we consider the probability that $n$ Regular pulsars and $m$ Irregular 
pulsars are observed, within totally $(n+m)$ pulsars.
Assuming that the $(n+m)$ sampling is random, this probability can be 
given by the bimodal distribution:
\begin{equation}
P_{n,m} = \frac{(n+m) !}{n!~m!} p_{\rm{r}}^n p_{\rm{ir}}^m .
\end{equation}
Hence, the probability that the observed number of Irregular pulsar is 
less than $M$ is given by the summation
\begin{equation}
P_{m < M} = \sum^M_{m=1} p_{N-m,m} ~,
\end{equation}
where $N = n+m$ is the total sample number.

\begin{figure}
  \centering
	\FigureFile(80mm,60mm){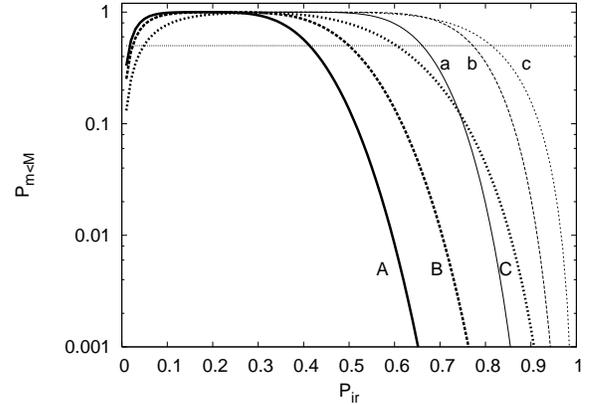}
  \caption{
    Probability of observing $M$ Irregular pulsars within $N$ samples,
    as the function of $p_{\rm{ir}}$.
    Thick lines (A, B and C) correspond to the XBPs of Irregular group, 
	while thin lines (a, b and c) show the Irregular and marginal LC pulsars.
	Curves with marks of (A, a), (B, b) and (C, c) denote results for 
    $L_{\rm X} > 5 \times 10^{36}$,  $L_{\rm X} > 1 \times 10^{37}$ and 
    $L_{\rm X} > 5 \times 10^{37}$, respectively.
  }
  \label{fig:p}
\end{figure}

In Fig.~{\ref{fig:p}}, we show the probability that we will observe $M$, 
or less, Irregular pulsars within the total sample $N$, as the function 
of $p_{\rm{ir}} = 2 \gamma / \pi$.
When we consider the bright pulsars with 
$L_{\rm X} > 5 \times 10^{36}$, $N = 40$ and $M = 16$.
This case is shown by thick solid line in the figure.
Imposing brighter criteria, $N = 29$ and $M = 14$ for 
$L_{\rm X} > 1 \times 10^{37}$, and 
$N = 14$ and $M = 8$ for $L_{\rm X} > 5 \times 10^{37}$, respectively.
Those results are shown by thick dashed and dotted lines in the figure.

Our diagnosis of light curves contains some uncertainties, 
and we denote the pulsars whose LC are marginally Regular by 
``Rm'' in the tables. 
If we count those marginal pulsars as Irregular members, the data count will 
be changed as $M = 26$ for $L_{\rm X} > 5 \times 10^{36}$, $M = 22$ for 
$L_{\rm X} > 1 \times 10^{37}$, and $M = 11$ for 
$L_{\rm X} > 5 \times 10^{37}$, respectively.
The probability functions for those cases are shown by thin lines in the same 
figure.

This figure shows that, in order to reconstruct the observed number count of 
Irregular pulsars, the probability $p_{\rm{ir}}$ should be small, under the 
present assumptions.
It directly means that the summation of the magnetic inclination angle 
$\theta_{\rm{m}}$ and the opening angles of beams $\varphi_{\rm{f}}$ and 
$\varphi_{\rm{p}}$ should be small.
In this analysis $\theta_{\rm{m}}$ and $\varphi$ are degenerated.
However, in order to explain the large number of bright Regular pulsars, anyway 
magnetic inclination $\theta_{\rm{m}}$ should be small.
In order to achieve $P_{m<M} > 0.5$, we need $p_{\rm{ir}} \lesssim 0.42$ 
($N=40$, $M=16$, $L_{\rm X} > 5 \times 10^{36}$) for the severest case 
(thick solid line in Fig.~\ref{fig:p}), and $p_{\rm{ir}} \lesssim 0.82$ 
($N=14$, $M=11$, $L_{\rm X} > 5 \times 10^{37}$, with "Rm"s)
for the lax case (thin dotted line in the figure).
On the other hand, $p_{\rm{ir}}$ should be larger than 0.02 -- 0.04, 
so highly collimated beams are also unacceptable.

About the magnetic inclination angles, Bulik et al. (2003) have analysed the 
ratio of single peak LCs over all the observed LCs, and they conclude that 
average $\theta_{\rm{m}}$ should be small in order to explain the observed 
ratio of single peak LC.
It is interesting and to be emphasised that, although our analysis is 
completely independent from the light curve analysis given by 
Bulik et al. (2003) which is using the pulse peak number,  
our result also suggests the same conclusion of small $\theta_{\rm{m}}$.

Another reason why the sample of Regular bright pulsars is over-sized might 
be the bias of observed energy bands.
In this analysis, we are assuming that the fan-beam and the pencil-beam 
will be emitted in different energy bands.
However, the energy bands in our sampling may not cover these different 
energy bands separately.
For instance, even if assume that, for a certain XBP, we have 
observational data in 2-4keV and 4-10keV, it is not always that the 
pencil-beam and the fan-beam could be observed separately in those 
two bands.
In our diagnosis, unfortunately the used energy bands are miscellaneous 
and are not standardized.
In actual diagnosis, 9 samples of 24 bright Regular pulsars are 
based on the observed LC in only 2 energy bands.
For the higher reliability, we clearly need further samples in broader 
energy bands from soft X-ray to gamma-ray.
For bright pulsars, the quality of the LC becomes better and small 
features in the LC can be distinguished, hence those bright pulsars 
may tend to be judged as Irregular type.
Also from this observational bias, this unexpectedly large fraction 
of Regular component looks a kind of new mystery of  XBP LCs.

In fact, the radiation process from the radiative column itself 
has been still unclear.
For the effective analysis, we need to know the exact energy band 
where the X-ray photons are emitted for both the pencil- and fan-beam. 
In order to know the reliable predominance of the emissions due to 
pencil-beam and fan-beam, we need to solve radiative-magneto-hydro 
equations.
Especially, from the point of view of beam opening angles, the radiation 
emitted from the radiative region in the strong magnetic field becomes 
important.

%-------------------------------------------------------------
%------------------------------------------------------------

\section{Summary and conclusion}

In this study, we analyze the observed light curves of X-ray binary pulsars.
We mainly focus on those XBP systems whose LCs are observed in multiple 
energy bands.
We found that some of them show different shapes of LC in different 
energy bands, and we named such LCs as "Irregular" type.
Looking for the origin of such an {\it irregularity}, we found that the X-ray 
luminosities of Irregular systems are higher than 
$\sim 5 \times 10^{36} [\rm{erg~s^{-1}}]$.
This criterion coincides with the critical limit for the existence of the 
accretion column with finite height, given by Basko, Sunyaev (1976). 
The coincidence of the X-ray luminosity limit of Irregular XBP
with the apparent condition of the accretion column is fairly consistent 
because of following reason:  
When the accretion column exists, the X-ray radiation will be emitted 
not only as the pencil-beam, but also as the fan-beam.
Hence, if we observe those different beams (and also combined beams) 
in different energy bands, the observed X-ray LC will be different in 
each energy band.

Additionally, we analysed the fraction of Irregular XBP in all the observed 
system.
It is apparent that not all the XBP with large $L_{\rm{X}}$ show 
Irregular LC.
This occurs because for the observation of Irregular LCs, the location of 
the observer has to satisfy the condition given in Eq.~(\ref{eq:incl}).
The probability to take such a location depends on the inclination 
angle of magnetic axis regarding to the rotational axis, 
and the opening angles of beams.
To explain the observational fraction of Irregular systems in our sample, 
the small inclination angle of the magnetic axis is strongly required.
This result is consistent with the independent analysis given by 
Bulik et al. (2003).
In our sampling, however, the completeness of the observed energy 
dependence is not so good.
In near future, we would like to collect larger number of samples with 
broad energy band, and analyze the data in detail.
At the same time, an improvement of the theoretical modelling of X-ray 
emission mechanisms in XBP is strongly required for better 
understanding of these systems.

\bigskip

The author would like to thank Michela Mapelli for 
her helpful discussions, careful reading, and comments.
He thanks Monica Colpi for her comments.
He thanks an anonymous referee for careful reading and 
valuable suggestions.

\begin{longtable}{l l c c c r r l}
\caption{List of X-ray pulsars in Galaxy}\label{tab:listG}
\hline\hline
No. & Name & LC & $P_{\rm s}$ & $P_{\rm o}$ & $L_{\rm x}$ & 
$B_{\rm p}$ & Comments \\
\endfirsthead
\hline\hline
No. & Name & LC & $P_{\rm s}$ & $P_{\rm o}$ & $L_{\rm x}$ & 
$B_{\rm p}$ & Comments \\
\endhead
\hline
\endfoot
\hline
\endlastfoot
G1  & XTEJ1808-369     & R   & 0.0025 & 0.084 & $3 \times 10^{36}$   & 
$< 1 \times 10^9$ & Poutanen et al. 2003 \\
G2  & 2EGJ0634+52      & --- & 0.0338 & 18    & $3.5 \times 10^{34}$ & 
$< 2 \times 10^9$ & Cusumano et al. 2000 \\
G3  & 1E1024.0-5732  & --- & 0.061  & ---   & $5.1 \times 10^{33}$ & ---    &  
wind?, Caraveo et al. 1989 \\
G4  & GROJ1744-28      & R   & 0.467  & 11.8  & $2 \times 10^{38}$   & 
$2 \times 10^{11}$  & Nishiuchi et al. 1999 \\
G5  & 2A1822-371     & Rm  & 0.592  & 0.23  & $2 \times 10^{37}$   & ---    & 
eclipse?, LMXB?, Parmar et al. 2000 \\
G6  & Her X-1      & IR  & 1.24   & 1.7   & $6 \times 10^{37}$   
& $2.9 \times 10^{12}$ & White et al. 1983, Nagase 1989 \\
G7  & 4U1901+03      & IR  & 2.736  & 22.6  & $8 \times 10^{37}$   
& $5 \times 10^{11}$  & wind?, Galloway et al. 2005 \\
G8  & 4U0115+634     & IR  & 3.6    & 24.3  & $3.0 \times 10^{37}$ 
& $1.3 \times 10^{12}$ & Santangelo et al. 1999 \\
G9  & EXO0332+53      & IR  & 4.37   & 34.3  & $1 \times 10^{38}$   
& $2.3 \times 10^{12}$ & Stella et al. 1985, Kreykenbohm et al. \\
    &              &     &        &       &        &        & 
2004, Tsygankov et al. 2005 \\ 
G10 & GROJ1750-27      & --- & 4.45   & 29.8  & $1 \times 10^{38}$ 
& ---    &  Scott et al. 1997 \\
G11 & AXJ1841.0-0535  & Rm  & 4.74   & 25    & $1.1 \times 10^{36}$ 
& ---    & 
Bamba et al. 2001 \\
G12 & Cen X-3      & Rm  & 4.84   & 2.1   & $1 \times 10^{38}$   & 
$2.2 \times 10^{12}$ & eclipse?, White et al. 1983, \\ 
    &              &     &        &       &        &        & 
Nagase 1989, Burderi et al. 2000 \\
G13 & RXJ1838-0301    & --- & 5.45   & ---   & $2 \times 10^{31}$   
& $1.6 \times 10^{14}$ & AXP?, Schwentker et al. 1994 \\
G14 & 1E1048.1-5937  & R   & 6.44   & 28    & $5 \times 10^{34}$   
& $1 \times 10^{14}$  &  AXP?, Tiengo et al. 2002, \\
   &              &     &        &       &        &        & Corbet et al. 1997 \\
G15 & 1E2259+586     & Rm  & 6.98   & ---   & $1 \times 10^{35}$   
& --- %%% $6 \times 10^{11}$  
&  Woods et al. 2004 \\
G16 & AXJ1845.0-0300   & R   & 7      & ---   & $1.6 \times 10^{35}$ 
& --- %%% $3 \times 10^{11}$  
& Torii et al. 1998 \\
G17 & 4U1626-673     & IR  & 7.6    & 0.02  & $2 \times 10^{37}$   
& $3.3 \times 10^{12}$ &  LMXB?, White et al. 1983, \\
    &              &     &        &       &        &        & 
Owens et al. 1997,\\
    &              &     &        &       &        &        & 
Chakrabarty et al. 1997 \\
G18 & 2S1553-54      & --- & 9.26   & 30    & $7 \times 10^{36}$   & ---    
& Kelley et al. 1983 \\
  
G19 & GS0834-430     & IR  & 12.3   & 105.8 & $1.1 \times 10^{37}$ & ---    
& Aoki et al. 1992, Wilson et al. 1997 \\
G20 & RXJ0648.1-4419  & --- & 13.2   & 1.55  & $4 \times 10^{32}$   
& $6 \times 10^{11}$  &  WD?, Israel et al. 1997 \\
G21 & XTEJ1946+274     & R   & 15.8   & 169.2 & $5.4 \times 10^{36}$ 
& $3.9 \times 10^{12}$ & Paul et al. 2001 \\
G22 & 2S1417-624     & R   & 17.6   & 42.1  & $8 \times 10^{36}$   & ---    & 
Finger et al. 1996 \\
G23 & GROJ1948+32      & Rm  & 18.7   & 40    & $2.1 \times 10^{37}$ 
& $2.5 \times 10^{13}$ & Chakrabarty et al. 1995, \\
   &              &     &        &       &        &        & Tsygankov \& Lutovinov 2005 \\
G24 & XTEJ1543-568     & R   & 27.1   & 75.6  & $1 \times 10^{37}$   & ---    & 
in't Zand et al. 2001b \\
G25 & GS1843+00      & IR  & 29.5   & ---   & $3 \times 10^{37}$   
& $1.7 \times 10^{12}$ & Nagase 1989 \\
G26 & RXJ0812.4-3114  & R   & 31.89  & 80    & $1.1 \times 10^{36}$ & ---    & 
eclipse?, Reig \& Rossi 1999\\
G27 & OAO1657-415     & Rm  & 37.7   & 10.4  & $1.6 \times 10^{37}$ 
& $3.2 \times 10^{12}$ & wind?, eclipse, White \& Pravdo 1979,  \\
   &              &     &        &       &        &        & White et al. 1983 \\
G28 & EXO2030+375     & IR  & 41.8   & 46.0  & $1 \times 10^{38}$   
& $1.1 \times 10^{13}$ & Nagase 1989 \\
G29 & 4U2135+57      & Rm  & 66.3   & ---   & $9.1 \times 10^{35}$ 
& $2.5 \times 10^{12}$ & Nagase 1989 \\
G30 & XTEJ1906+09      & IR  & 89.2   & ---   & $2 \times 10^{35}$   & ---    & 
wind?, Marsden et al. 1998 \\
G31 & GROJ1008-57      & IR  & 93.5   & 248   & $4.1 \times 10^{36}$ 
& $7.6 \times 10^{12}$ & Shrader et al. 1999 \\
G32 & 4U1850-03      & R   & 94.8   & 242.2 & $6 \times 10^{36}$   & ---    & 
Nagase 1989 \\
G33 & 4U0728-260     & Rm  & 103.2  & 34.5  & $2.8 \times 10^{35}$ & ---    & 
Corbet \& Peele 1997 \\
G34 & A0535+262     & IR  & 105    & 110   & $9 \times 10^{37}$   
& $4.3 \times 10^{12}$ & eclipse?, Orlandini et al. 2004, \\
   &              &     &        &       &        &        & Bildsten et al. 1997 \\
G35 & AXJ1838-0655    & Rm  & 111    & ---   & $3.2 \times 10^{35}$ & ---    & 
Koyama et al. 1991 \\
G36 & GX4+1     & Rm  & 115    & 304   & $1 \times 10^{38}$   & ---    & 
White et al. 1983, Nagase 1989 \\
G37 & SAXJ1802.7-201   & R   & 139.5  & 4.6   & $2 \times 10^{36}$   & ---    & 
eclipse?, wind?, Augello et al. 2003 \\
G38 & AXJ1820.5-1434  & R   & 152.3  & ---   & $9 \times 10^{34}$   & ---    & 
Kinugasa et al. 1998 \\
G39 & 1SAXJ1324.4-6200  & Rm  & 170    &$>$100 & $1.1 \times 10^{34}$ 
& $3 \times 10^{12}$  & Angelini et al. 1998 \\
G40 & GROJ2058+42      & IR  & 198    & 110   & $1 \times 10^{38}$   & ---    & 
Wilson et al. 1999 \\
G41 & XTEJ1858+034     & R   & 202.7  & ---   & $9 \times 10^{34}$   & ---    & 
Paul \& Rao 1998 \\
G42 & RXJ0440.9+4431  & R   & 203    & ---   & $3 \times 10^{34}$   & ---    & 
Reig \& Roche 1999 \\
G43 & AXJ1749.2-2725  & Rm  & 220.4  & ---   & $2.6 \times 10^{35}$ & ---    & 
Torii et al. 1998\\
G44 & GX304-1	   & Rm  & 272    & 132   & $1 \times 10^{36}$   & ---    & 
White et al. 1983 \\
G45 & Vela X-1     & IR  & 283    & 8.96  & $6 \times 10^{36}$   
& $2.1 \times 10^{12}$ & wind?, White et al. 1983, \\
   &              &     &        &       &        &        & Nagase et al. 1989 \\
G46 & 4U1145-61      & Rm  & 292    & 187.5 & $7.4 \times 10^{34}$ & ---    & 
White et al. 1983\\
G47 & 1E1145.1-6141  & R   & 298    & 5.65  & ---    & ---    & 
Ray \& Chakrabarty 2002, \\
   &              &     &        &       &        &        & White et al. 1980 \\
G48 & SAXJ2103.5+4545  & R   & 358.6  & 12.7  & $3 \times 10^{36}$   
& $1.6 \times 10^{12}$ & Hulleman et al. 1998, Sidoli et al. 2005 \\
G49 & XTEJ1855-026     & R   & 361.1  & 6.1   & $2 \times 10^{36}$   & ---    & 
wind?, Corbet et al. 1998 \\
G50 & 4U2206+543     & --- & 392    & 9.6   & $2.5 \times 10^{35}$ & ---    & 
2 companion?, Corbet \& Peele 2001 \\
G51 & A1118-615     & R   & 405    & ---   & $5 \times 10^{36}$   & ---    & 
Coe et al. 1994 \\
G52 & GPS1722-363     & R   & 413.7  & ---   & $9 \times 10^{36}$   & ---    & 
wind?, Nagase 1989, \\
   &              &     &        &       &        &        & Zurita Haras et al. 2005 \\ 
G53 & 1SAXJ1452.8-5949  & --- & 437    & ---   & $8.7 \times 10^{33}$ & ---    & 
Oosterbroke et al. 1999 \\
G54 & 4U1907+09      & IR  & 438    & 8.38  & $4 \times 10^{37}$   
& $1.6 \times 10^{12}$ & Makishima et al. 1984, \\
   &              &     &        &       &        &        & in't Zand et al. 1998 \\
G55 & 4U1538-522     & R   & 529    & 3.73  & $4 \times 10^{36}$   
& $1.8 \times 10^{12}$ & wind?, White et al. 1983, \\
   &              &     &        &       &        &        & Robba et al. 2001 \\
G56 & 4U1909+07      & --- & 605    & ---   & $6 \times 10^{36}$   & ---    & 
wind?, Lavine et al. 2004 \\
G57 & GX301-2      & Rm  & 695    & 41.5  & $6 \times 10^{36}$   
& $3.1 \times 10^{12}$ & wind?, Koh et al. 1997,  \\
   &              &     &        &       &        &        & White et al. 1983 \\
G58 & 2RXSJ130159-6358  & R   & 700    & ---   & $1 \times 10^{35}$   
& ---    & Chenyakova et al. 2005 \\
G59 & AXJ170006-4157  & ---  & 714.5  & ---   & $7 \times 10^{34}$   
& ---    & Torii et al. 1999 \\
G60 & 1WGAJ1958.2-3232  & Rm  & 733    & 0.2   & $6 \times 10^{32}$   
& ---    & Norton et al. 2002 \\
G61 & X Per        & R   & 837    & 250   & $3 \times 10^{35}$   
& $3.3 \times 10^{12}$ & wind?, White et al. 1983, \\
   &              &     &        &       &        &        & Di Salvo et al. 1998 \\
G62 & RXJ1037.5-5647  & R   & 862    & ---   & $4.5 \times 10^{35}$ & ---    & 
Reig \& Roche 1999\\
G63 & IGRJ16393-4643   & R   & 912    & ---   & $1.2 \times 10^{36}$ & ---    & 
wind?, Bodaghee et al. 2005 \\
G64 & SAXJ2239.3+6116  & R   & 1247   & 262.6 & $2.3 \times 10^{36}$ & ---    & 
in't Zand et al. 2001a \\
G65 & AXJ1631.9-4752  & R   & 1309   & ---   & $1.2 \times 10^{36}$ & ---    & 
Lutovinov et al. 2005, \\
   &              &     &        &       &        &        & Rodriguez et al. 2003, 2005 \\
G66 & RXJ0146.9+6121  & R   & 1413   & ---   & $3.5 \times 10^{35}$ & ---    & 
Mereghetti et al. 2000 \\
G67 & 2S0114+650     & R   & 10008  & 11.6  & $1.1 \times 10^{36}$ 
& $2.5 \times 10^{12}$ & wind?, Hall et al. 2000, \\
   &              &     &        &       &        &        & Bonning \& Falanga 2005 \\
G68 & 1RXJ170849.0-400910 & Rm & 11 & --- & $2 \times 10^{36}$ & $1.6 \times 10^{15}$ 
&  certain AXP candidate, not included \\ 
    &              &     &        &       &        &        & 
in the statistics, Sugizaki et al. 1997, \\
    &              &     &        &       &        &        & 
Campana et al. 2007,  \\   
G69 & X1839-04      & --- & 81.1   & ---   & ---    & ---    & No available LC \\
G70 & 1239-599     & --- & 191    & ---   & ---    & ---    & No available LC \\
\hline
\end{longtable}

\begin{longtable}{l l c c c r r l}
  \caption{List of X-ray pulsars in SMC}\label{tab:listS}
  \hline\hline
No. & Name & LC & $P_{\rm s}$ & $P_{\rm o}$ & $L_{\rm x}$ & 
$B_{\rm p}$ & Comments \\
\endfirsthead
  \hline\hline
No. & Name & LC & $P_{\rm s}$ & $P_{\rm o}$ & $L_{\rm x}$ & 
$B_{\rm p}$ & Comments \\
\endhead
  \hline
\endfoot
  \hline
\endlastfoot
S1 & SMCX-1        & Rm  & 0.71  & ---  & $3.8 \times 10^{38}$ & --- & 
Yokogawa et al. 2000b \\
S2 & SMCX-2        & IR  & 2.37  & ---  & $4.7 \times 10^{38}$ & --- & 
Yokogawa et al. 2001c, Corbet et al. 2001 \\	
S3 & RXJ0059.2-7138   & R   & 2.76  & ---  & $5.0 \times 10^{37}$ & --- & 
Kohno et al. 2000, Huges et al. 1994\\
S4 & AXJ0105-722       & Rm  & 3.34  & ---  & $1.5 \times 10^{35}$ & --- & 
Yokogawa et al. 2000b \\
S5 & XTEJ0052-723      & --- & 4.78  & ---  & $7.2 \times 10^{37}$ & --- & 
Laycock et al. 2003 \\
S6 & CXOUJ011004-72113  & --- & 5.4   & ---  & $1 \times 10^{35}$   & --- & 
Lamb et al. 2002 \\
S7 & SMCX-3        & --- & 7.78  & 44.6	& $6 \times 10^{37}$	 & --- & 
Edge et al. 2004 \\
S8 & RXJ0051.8-7231   & R   & 8.9   & 185  & $1.4 \times 10^{36}$ & --- & 
Israel et al. 1997 \\
S9 & AXJ0049-732      & --- & 9.13  & 91.5	& $4.1 \times 10^{35}$ & --- & 
AXP?, Ueno et al. 2000 \\
S10 & RXJ0052.1-7319  & R   & 15.3  & ---  & $1.5 \times 10^{37}$ & --- & 
Finger et al. 2001, Kahabka et al. 2000 \\	
S11 & RXJ0050.9-7310  & --- & 16.6  & 189  & $3.7 \times 10^{36}$ & --- & 
Lamb et al. 2002 \\
S12 & RXJ0117.6-7330  & IR  & 22.07 & ---  & $1.2 \times 10^{38}$ & --- & 
Macomb et al. 1999 \\
S13 & XTEJ0050-732\#2  & --- & 25.5  & ---  & $3.0 \times 10^{36}$ & --- & 
Lamb et al. 2002 \\
S14 & XTEJ0111.2-7317  & R   & 31.03 & ---  & $2.0 \times 10^{38}$ & --- & 
SNR?, Yokogawa et al. 2000a \\
S15 & J0055.4-7210  & --- & 34.08 & ---  & $1.1 \times 10^{35}$ & --- & 
Edge et al. 2004 \\
S16 & XTEJ0053-724     & R   & 46.63 & 139  & $7.4 \times 10^{36}$ & --- & 
Yokogawa et al. 2000b \\
S17 & XTEJ0055-724     & Rm  & 59    & 65   & $4.3 \times 10^{37}$ & --- & 
Santangelo et al. 1998, Sasaki et al. 2003 \\
S18 & AXJ0049-729     & R   & 74.67 & 33   & $7.5 \times 10^{36}$ & --- & 
Yokogawa et al. 1999 \\
S19 & XTEJ0052-725     & R   & 82.46 & ---  & $3.4 \times 10^{36}$ & --- & 
Edge et al. 2004 \\
S20 & AXJ0051-722     & R   & 91.1  & 88.4 & $2.9 \times 10^{37}$ & --- & 
Yokogawa et al. 2000b \\
S21 & SMC95        & Rm  & 95    & ---  & $2 \times 10^{37}$   & --- & 
Laycock et al. 2002 \\
S22 & AXJ0057.4-7325  & Rm  & 101.5 & ---  & $1.2 \times 10^{36}$ & --- & 
Yokogawa et al. 2000e \\
S23 & RXJ0053.5-7227  & --- & 138   & 125  & $1.2 \times 10^{35}$ & --- & 
Edge et al. 2004 \\
S24 & 2E0054.4-7237  & Rm  & 140.1 & ---  & $4.0 \times 10^{34}$ & --- & 
Sasaki et al. 2003 \\
S25 & RXJ0057.8-7207  & Rm  & 152.3 & ---  & $4.3 \times 10^{35}$ & --- & 
Sasaki et al. 2003 \\
S26 & RXJ0052.9-7158  & R   & 167.8 & ---  & $2.0 \times 10^{37}$ & --- & 
Yokogawa et al. 2001b \\
S27 & RXJ0051.9-7311  & Rm  & 172.4 & 147  & $5.6 \times 10^{36}$ & --- & 
Yokogawa et al. 2000d \\
S28 & RXJ0059.3-7223  & IR  & 202   & ---  & $3.2 \times 10^{35}$ & --- & 
wind?, Majid et al. 2004 \\
S29 & RXJ0047.3-7312  & Rm  & 263   & 49   & $1.8 \times 10^{36}$ & --- & 
Majid et al. 2004 \\
S30 & AXJ0058-720     & R   & 280.4 & ---  & $1.6 \times 10^{36}$ & --- & 
Sasaki et al. 2003, Tsujimoto et al. 1999 \\
S31 & RXJ0101.0-7206  & --- & 304   & ---  & $1.3 \times 10^{36}$ & --- & 
Macomb et al. 2003 \\
S32 & AXJ0051-733     & Rm  & 323.2 & 1.4  & $1.8 \times 10^{36}$ & --- &  
Imanishi et al. 1999, Yokogawa et al. 2000b \\
S33 & SAXJ0103.2-7209  & R   & 345.2 & ---  & $1.5 \times 10^{36}$ & --- & 
Sasaki et al. 2003, Israel et al. 2000, \\
    &              &     &       &      &        &     &  
Yokogawa et al. 2000b \\
S34 & RXJ0101.3-7211  & R   & 455   & ---  & $7.3 \times 10^{35}$ & --- & 
Sasaki et al. 2001, 2003\\
S35 & AXJ0054.8-7244  & Rm  & 500.0 & 268  & $5.5 \times 10^{35}$ & --- & 
Haberl et al 2004 \\
S36 & CXOUJ005736-7219  & --- & 565   & 95.3 & $1.2 \times 10^{36}$ & --- & 
Macomb et al. 2003, Edge et al. 2004 \\
S37 & SXP701       & R   & 702   & 413  & $4 \times 10^{35}$   & --- & 
Haberl et al. 2004 \\
S38 & RXJ0049.7-7323  & Rm  & 755.5 & 394  & $7.7 \times 10^{35}$ & --- & 
Yokogawa et al. 2000c \\	
S39 & RXJ0103.6-7201  & IR  & 1323  & ---  & $5.4 \times 10^{36}$ & --- & 
Haberl \& Pietsch 2005 \\
S40 & XTEJ0055-727     & --- & 18.4  & 34.8 & $2.6 \times 10^{37}$ & --- 
& No available LC \\
S41 & AXJ0043-737     & --- & 87.6  & ---  & ---    & --- & No available LC \\
\hline
\end{longtable}

\begin{longtable}{l l c c c r r l}
  \caption{List of X-ray pulsars in LMC}\label{tab:listL}
  \hline\hline
No. & Name & LC & $P_{\rm s}$ & $P_{\rm o}$ & $L_{\rm x}$ & 
$B_{\rm p}$ & Comments \\
\endfirsthead
  \hline\hline
No. & Name & LC & $P_{\rm s}$ & $P_{\rm o}$ & $L_{\rm x}$ & 
$B_{\rm p}$ & Comments \\
\endhead
  \hline
\endfoot
  \hline
\endlastfoot
L1  & RXJ0502.9-6626  & --- & 4.1   & ---  & $4 \times 10^{37}$   & --- & 
Schmidtke et al. 1995 \\	
L2  & LMC X-4      & IR  & 13.7  & 24.5 & $1 \times 10^{37}$   & --- & 
Woo et al. 1996, Paul et al. 2002, \\	
    &              &     &        &       &        &        & 
Haberl et al. 2003 \\
L3  & RXJ0529.8-6556  & Rm  & 69.5  & ---  & $2 \times 10^{36}$   & --- & 
Haberl et al. 1997, 2003 \\
L4  & SAXJ0544.1-710   & Rm  & 96.08 & ---  & $2 \times 10^{36}$   & --- & 
Cusumano et al. 1998 \\
L5  & EXO054011-65512 & --- & 272   & ---  & $3.9 \times 10^{35}$ & --- & 
Haberl et al. 2003 \\
L6  & A0535-668     & --- & 0.068 & 16.7 & $1 \times 10^{39}$   & --- & 
Skinner et al. 1982 \\
\end{longtable}

\end{document}